\documentclass[pre,preprint,longbibliography]{revtex4-1}%
\usepackage{amssymb}
\usepackage{amsfonts}
\usepackage{amsmath}
\usepackage{xcolor} 
\usepackage{ulem} 
\usepackage{graphicx}%
\providecommand{\U}[1]{\protect\rule{.1in}{.1in}}
\begin{document}
\title{Classical Density Functional Theory applied to the solid state}
\author{James F. Lutsko}
\homepage{http://www.lutsko.com}
\email{jlutsko@ulb.ac.be}
\author{C\'{e}dric Schoonen}
\affiliation{Center for Nonlinear Phenomena and Complex Systems CP 231, Universit\'{e}
Libre de Bruxelles, Blvd. du Triomphe, 1050 Brussels, Belgium}

\begin{abstract}
The standard model of classical Density Functional Theory for pair potentials consists of a hard-sphere functional plus a mean-field term accounting for long ranged attraction. However, most implementations using sophisticated Fundamental Measure hard-sphere functionals suffer from potential numerical instabilities either due to possible instabilities in the functionals themselves or due to implementations that mix real- and Fourier-space components inconsistently. Here, we present a new implementation based on a demonstrably stable hard-sphere functional that is implemented in a completely consistent manner. The present work does not depend on approximate spherical integration schemes and so is much more robust than previous algorithms. The methods are illustrated by calculating phase diagrams for the solid state using the standard Lennard-Jones potential as well as a new class of potentials recently proposed by Wang et al (Phys. Chem. Chem. Phys. 22, 10624 (2020)). The latter span the range from potentials for small molecules to those appropriate to colloidal systems simply by varying a parameter. We verify that cDFT is able to semi-quantitatively reproduce the phase diagram in all cases. We also show that for these problems computationally cheap Gaussian approximations are nearly as good as full minimization based on finite differences. 
\end{abstract}

\date{\today }
\maketitle

\section{Introduction}

Classical Density Functional Theory (cDFT) is an exact theory that has become
a versatile tool for the studying the properties of inhomogeneous systems such
as fluid interfaces and solids at length scales going down to the molecular
level\cite{Evans1979, lutsko:acp}. Conceptually, cDFT calculations involve the
minimization of a functional of the local density resulting in both the
equilibrium density distribution and the free energy of the system. Notable
recent applications include the description of wetting
phenomena\cite{Evans23901}, the calculation of hydration free-energies and
microscopic structure of molecular solutes\cite{Borgis} and the description of
crystallization pathways\cite{Lutsko_HCF}. An important part of this utility
lies in the highly-developed description of correlations due to
excluded-volume effects that are arise whenever molecules interact via
potentials having divergent short-ranged repulsion as is the case, e.g., in
simple fluids and is captured in such commonly used models as the
Lennard-Jones, Stillinger-Weber and hard-core Yukawa potentials. This
capability is most highly developed in models based on Fundamental Measure
Theory which was first introduced by Rosenfeld\cite{rosenfeld1}, inspired by
exact results due to Percus\cite{Percus1,Percus2,Percus3}, and that has been
steadily developed over the last 30 years (see, e.g., Ref. \onlinecite{Roth_2010}).

Despite this progress, several issues have slowed the application of cDFT to
the most challenging of inhomogeneous systems, namely the solid phase where
the local density varies by many orders of magnitude over distances of the
molecular diameter. Indeed, the application to solids is sufficiently
challenging that until recently such applications as existed were based on
simplifications such as the modeling of the local density as a sum of Gaussian
profiles. The application of the full machinery of cDFT to the solid phase
only began with the work of Oettel and coworkers in 2010\cite{Oettel2010} and there remain
relatively few applications. Perhaps the main reason for this is that these
early calculations seemed to be very delicate and plagued by numerical
instabilities. This has led, e.g., to calculations only being possible at fixed particle number rather than at fixed chemical potential as is more natural in cDFT(see, e.g. Ref. \cite{Martin}).  Recently, it has been proposed that these difficulties are
traceable to instabilities inherent in some of the most popular models (based
on the so-called White Bear FMT\ functionals) as well as to certain
inconsistencies in the implementation of the calculations (as explained in
more detail below)\cite{LutskoLam, esFMT}. The solution to the first problem is the careful selection
of models that are provably stable which, while more heuristic than the most
advanced models, are sufficiently accurate for many purposes. One of the main
purposes of the present work is to present a solution to the second
problem:\ namely, a robust implementation that is free of instabilities. 

The second goal of this work is to examine the transferability of the cDFT
model - i.e. its robustness when applied to different interaction potentials.
The recent work on solids has mostly focused on the Lennard-Jones potential
and it relatively good results have been reported, compared to simulation.
Here, we also examine a new class of potentials introduced by Wang et al\cite{Wang} who
thoroughly characterized their vapor-liquid-solid phase diagrams and so
provide an excellent test case. We focus on the two examples studied in detail
in that work:\ namely, an analog of the Lennard-Jones potential and a second
that models colloids. Colloids and simple fluids have qualitatively different
phase diagrams and it is therefore of interest to verify to what extent cDFT
is able to describe such variations.

In the following, we first review the basic elements of the standard cDFT
model consisting of the sum of an ideal gas contribution, an\ FMT hard-sphere
contribution and a mean-field term that captures the details of the potential.
We also present our numerical implementation of this model. In the next
Section, we present our results first for the the Lennard-Jones and
WHDF\ potential parameterized as described above for both simple fluids and
colloids. We have calculated the vapor-liquid-solid phase diagrams, the
liquid-vapor surface tensions and we also present some details of the solid
phase such as the asymmetry of the density distributions,the vacancy
concentration and the difference between the HCP and FCC phases. We end with some conclusions.

\section{Theory}

\subsection{The standard cDFT model}

The fundamental quantity in cDFT is the local density $\rho\left(
\mathbf{r}\right)  $. It is important to emphasize that for an equilibrium
system this is identical to the as the one-body probability distribution and
as such is a microscopic quantity that involves no coarse-graining or other
approximations and, as such, it gives a description of the equilibrium density
distribution valid down to the smallest length scales. In cDFT, the local
density of an equilibrium system is determined by minimizing a functional
denoted here as $\Lambda\left[  \rho\right]  $. The formal theory underlying
cDFT assures us that such a functional exists, that it is unique and that when
evaluated at the equilibrium density, it is equal to the grand-canonical free
energy of the system, $\Omega$. The "standard" cDFT model is written as
\begin{equation}
\Lambda\left[  \rho\right]  =F^{\left(  \text{id}\right)  }\left[
\rho\right]  +F^{\left(  \text{HS}\right)  }\left[  \rho\right]  +\frac{1}%
{2}\int\rho\left(  \mathbf{r}_{1}\right)  \rho\left(  \mathbf{r}_{2}\right)
w_{att}\left(  \mathbf{r}_{1},\mathbf{r}_{2}\right)  d\mathbf{r}_{1}d\mathbf{r}%
_{2}+\int\rho\left(  \mathbf{r}\right)  \left(  \phi\left(  \mathbf{r}\right)
-\widetilde{\mu}\right)  d\mathbf{r}%
\end{equation}
where the terms are, in sequence, the ideal gas contribution, the hard-sphere
contribution, the mean-field contribution and the external field contribution.
The ideal-gas part is
\begin{equation}
F^{\left(  \text{id}\right)  }\left[  \rho\right]  =k_{B}T\int\left(
\rho\left(  \mathbf{r}\right)  \ln l^{3}\rho\left(  \mathbf{r}\right)
-\rho\left(  \mathbf{r}\right)  \right)  d\mathbf{r}%
\end{equation}
where $k_{B}$ is Boltzmann's constant, $T$ is the temperature and $l$ is any
convenient length scale. The next two terms depend on the intermolecular
potential which we take to be a pair potential $v\left(  r\right)  $. This is
separated into a repulsive part, $v_{0}\left(  r\right)  $ and an attractive
part $w_{att}\left(  r\right)  =v\left(  r\right)  -v_{0}\left(  r\right)  $. We use
the WCA\ prescription whereby $v_{0}\left(  r\right)  =v\left(  r\right)
-v\left(  r_{0}\right)  $ for $r<r_{0}$ and zero for $r\geq r_{0}$ where
$r_{0}$ is the minimum of the potential. Then, an effective hard-sphere
diameter $d$ is constructed using any convenient prescription:\ here, we use
the Barker-Henderson recipe%
\begin{equation}
d=\int_{0}^{r_{0}}\left(  1-e^{-\beta v_{0}\left(  r\right)  }\right)  dr
\end{equation}
where $\beta\equiv1/k_{B}T$. For the hard-sphere contribution, we use the FMT
ansatz
\begin{equation}
F^{\left(  \text{HS}\right)  }\left[  \rho\right]  =k_{B}T\int\Phi\left(
\eta\left(  \mathbf{r}\right)  ,s\left(  \mathbf{r}\right)  ,\mathbf{v}\left(
\mathbf{r}\right)  \right)  d\mathbf{r}%
\end{equation}
where the fundamental measures are
\begin{equation}
n^{\left(  \alpha\right)  }\left(  \mathbf{r}\right)  =\int w^{\left(
\alpha\right)  }\left(  \mathbf{r-r}^{\prime}\right)  \rho\left(
\mathbf{r}^{\prime}\right)  d\mathbf{r}^{\prime}%
\end{equation}
with the hard-sphere radius $R=d/2$ and the weights $w^{\left(  \eta\right)
}\left(  \mathbf{r}\right)  =\Theta\left(  R-r\right)  $, $w^{\left(
s\right)  }\left(  \mathbf{r}\right)  =\delta\left(  R-r\right)  $ and
$w^{\left(  \mathbf{v}\right)  }\left(  \mathbf{r}\right)  =\frac{\mathbf{r}%
}{r}\delta\left(  R-r\right)  $ where $\Theta\left(  x\right)  $ is the step
function equal to $1$ if $x>0$ and zero otherwise. Different FMT models are
distinguished by the form of the function $\Phi\left(  \eta,s,\mathbf{v}%
\right)  $ and here we use the modified RSLT function%
\begin{equation}
\Phi\left(  \eta,s,\mathbf{v}\right)  = - \frac{1}{\pi\sigma^{2}} s \ln
(1-\eta) + \frac{1}{2 \pi\sigma} \frac{s^{2} - v^{2}}{1-\eta} +\frac{1}{24
\pi} \frac{s^{3}\left(  1-(v^{2}/s^{2})\right)  ^{3}}{(1-\eta)^{2}} \phi
_{2}(\eta)
\end{equation}
with
\begin{equation}
\phi_{2}(\eta) = 1-\frac{-2\eta+ 3\eta^{2} -2(1-\eta)^{2} \ln(1-\eta)}%
{3\eta^{2}}%
\end{equation}
which is easily shown to be free of instabilities\cite{LutskoLam}. The
mean-field term is already given explicitly and we just note that $w_{att}\left(
\mathbf{r}_{1},\mathbf{r}_{2}\right)  =w_{att}\left(  \left\vert \mathbf{r}%
_{1}-\mathbf{r}_{2}\right\vert \right)  $, the attractive part of the
potential. Finally, $\phi\left(  \mathbf{r}\right)  $ represents any external
one-body field (this plays no role in the present work) and $\widetilde{\mu
}=\mu-\ln\left(  \frac{\Lambda}{l}\right)  ^{3}$ where $\mu$ is the chemical
potential and $\Lambda$ is the thermal wavelength. The model functional has
been presented for a single component system but generalization to multiple
components is straightforward.

\subsection{Implementation}

We discretize the density on a cubic lattice of $N_x \times N_y \times N_z$ points with spacing  $\Delta$ and
our key approximation is that the density field is approximated by trilinear
interpolation of the values of the density at the lattice positions. So,
working in units of $\Delta=1$, and defining $\rho_{\mathbf{S}}=\rho\left(
S_{x}\widehat{\mathbf{x}}+S_{y}\widehat{\mathbf{y}}+S_{z}\widehat{\mathbf{z}%
}\right)  $ where $S_{x},S_{y},S_{z}$ are integers with $1\leq S_{x}\leq
N_{x}$, etc. we write
\begin{equation}
\rho\left(  \mathbf{r}\right)  =\sum_{\mathbf{I}=0,1}A_{\mathbf{I}}\left(
\mathbf{r-S}\left(  \mathbf{r}\right)  \right)  \rho\left(  \mathbf{S}\left(
\mathbf{r}\right)  +\mathbf{I}\right)
\end{equation}
where $\mathbf{S}\left(  \mathbf{r}\right)  $ is the lattice point nearest the
origin which is a corner of the computational cell containing the point
$\mathbf{r}$, i.e. that for which $\left\vert S_{x}\right\vert \leq\left\vert
r_{x}\right\vert <\left\vert S_{x}\right\vert +1$, etc and $\sum
_{\mathbf{I}=0,1}$ is shorthand for $\sum_{I_{x}=0,1}\sum_{I_{y}=0,1}%
\sum_{I_{z}=0,1}$. The coefficients are
\begin{equation}
A_{\mathbf{I}}\left(  \mathbf{r}\right)  =\left(  \delta_{I_{x}0}\left(
1-2x\right)  +x\right)  \left(  \delta_{I_{y}0}\left(  1-2y\right)  +y\right)
\left(  \delta_{I_{z}0}\left(  1-2z\right)  +z\right)
\end{equation}
which just means that we interpolate the density linearly between the lattice positions forming the cell containing the point $\mathbf{r}$.

\subsubsection{Evaluation of the ideal part of the free energy}

In principle, one would like to write the exact expression for the ideal gas
contribution as
\begin{equation}
F^{\left(  \text{id}\right)  }\left[  \rho\right]  =\sum_{\mathbf{S}}%
\int_{\mathcal{C}\left(  \mathbf{S}\right)  }\left(  \rho\left(
\mathbf{r}\right)  \ln l^{3}\rho\left(  \mathbf{r}\right)  -\rho\left(
\mathbf{r}\right)  \right)  d\mathbf{r}%
\end{equation}
where $\mathcal{C}\left(  \mathbf{S}\right)  $ is the cell for which the
corner at lattice site $\mathbf{S}$ is the closest to the origin. One could
then, in each of the integrals insert the trilinear interpolation and perform
the integral. However, given the non-linearity of the expression, we have found
this prohibitively difficult to do analytically, although it seems in
principle possible. We have therefore used the simpler approximation whereby
the density is treated as being constant in each cell giving the
straightforward discretization%
\begin{equation}
F^{\left(  \text{id}\right)  }\left[  \rho\right]  \simeq k_{B}T\Delta^{3}%
\sum_{\mathbf{S}}\left(  \rho_{\mathbf{S}}\ln l^{3}\rho_{\mathbf{S}}%
-\rho_{\mathbf{S}}\right)  .
\end{equation}
Our benchmarking suggests that this remains surprisingly accurate even for
relatively highly localized density distributions (see Supplementary
material). Note that the trilinear interpolation, in any case, gives the exact
result for the average number of particles
\begin{equation}
\left\langle N\right\rangle =\int\rho\left(  \mathbf{r}\right)  d\mathbf{r=}%
\sum_{\mathbf{S}}\int_{\mathcal{C}\left(  \mathbf{S}\right)  }\rho\left(
\mathbf{r}\right)  d\mathbf{r}\underset{\text{trilinear}}{\mathbf{=}}%
\Delta^{3}\sum_{\mathbf{S}}\rho_{\mathbf{S}}.
\end{equation}

\subsubsection{Evaluation of hard-sphere contribution}

The FMT contribution to the free energy is discretized in the simplest way as
\begin{equation}
F^{\left(  \text{HS}\right)  }\left[  \rho\right]  \simeq k_{B}T\Delta^{3}%
\sum_{\mathbf{S}}\Phi\left(  \eta_{\mathbf{S}},s_{\mathbf{S}},\mathbf{v}%
_{\mathbf{S}}\right)  .
\end{equation}
In principle, one could use a more accurate scheme but since the fundamental
measures are more slowly varying than the density itself, this seems
sufficiently accurate on a computational lattice that is fine enough to
resolve the density. The main effort is therefore the evaluation of the
fundamental measures at the lattice points.

The standard method of computing the fundamental measures is to note that they
are convolutions so that they can be efficiently evaluated by transforming the
weights and density to Fourier space, multiplying and then performing and
inverse Fourier transform. The obvious strategy is to Fourier transform the
weights analytically since they are simple to do while the density requires a
discrete Fourier transform. This in principle leads to an inconsistency where
it is impossible to prove that the resulting approximation preserves the
stability of the free energy functional. We therefore follow the example of
Ref.\onlinecite{LutskoLam} and do everything consistently on the lattice. This
leads to the practical problem that the FMT weights are not well adapted to
evaluation on a lattice: the step function can be handled in obvious ways but
the delta-functions are less obvious. In Ref.\onlinecite{LutskoLam} this was
addressed by using pre-compiled integration points on a sphere but this is a
less than optimal solution as it leads to small-scale variations due to the
pseudo-random nature of the points. Here, we present a straightforward
alternative that avoids these issues by performing the necessary integrals
analytically giving a fast and easily coded implementation.

The fundamental measures evaluated at the lattice sites are
\begin{equation}
n^{\left(  \alpha\right)  }\left(  \mathbf{S}\right)  =\int w^{\left(
\alpha\right)  }\left(  \mathbf{S-r}\right)  \rho\left(  \mathbf{r}\right)
d\mathbf{r} \label{fmt}%
\end{equation}
so that substituting the trilinear interpolation for the density and making a
few simple transformations leads to
\begin{equation}
n^{\left(  \alpha\right)  }\left(  \mathbf{S}\right)  =\sum_{\mathbf{S}%
^{\prime}}\widetilde{w}^{\left(  \alpha\right)  }\left(  \mathbf{S-S}^{\prime
}\right)  \rho\left(  \mathbf{S}^{\prime}\right)
\end{equation}
with%
\begin{equation}
\widetilde{w}^{\left(  \alpha\right)  }\left(  \mathbf{S}\right)
=\sum_{\mathbf{K}=0,1}\int_{\mathcal{C}\left(  0\right)  }w^{\left(
\alpha\right)  }\left(  \mathbf{S+K-r}\right)  A_{\mathbf{K}}\left(
\mathbf{r}\right)  d\mathbf{r}%
\end{equation}
where $\mathcal{C}\left(  0\right)  $ is the cubic cell defined by the
diagonal points $(0,0,0)$ and $(1,1,1)$. These integrals can in fact be
performed numerically cell-by-cell in the case of the step-function weight but the
delta-functions still pose problems. Fortunately, it turns out that these
expressions can, with some effort, be evaluated analytically for the simple
weights used in FMT. Details and explicit expressions are given in Appendix
\ref{AppW} and in the Supplementary text. With the real-space weights in hand, the evaluation of the
fundamental proceeds by means of discrete Fourier transform in the obvious way.

\subsubsection{The mean-field contribution}

Evaluation of the mean-field term using the trilinear interpolation scheme for the density results in 
\begin{equation}
F^{\left(  \text{MF}\right)  }\left[  \rho\right]  =\frac{1}{2}\Delta^{6}%
\sum_{\mathbf{S,S}^{\prime}}\rho_{\mathbf{S}}\rho_{\mathbf{S}^{\prime}%
}\widetilde{w}^{\left(  \text{MF}\right) }_{\mathbf{S-S}^{\prime}}%
\end{equation}
where the weights have the form%
\begin{equation}
\widetilde{w}^{\left(  \text{MF}\right) }_{\mathbf{S}}=\Delta^{-6}\sum_{\mathbf{L}=0,1}\sum_{\mathbf{K}%
=0,1}\int_{\mathcal{C}\left(  \mathbf{0}\right)  }\int_{\mathcal{C}\left(
\mathbf{0}\right)  }A_{\mathbf{L}}\left(  \mathbf{r}_{1}\right)
A_{\mathbf{K}}\left(  \mathbf{r}_{2}\right)  w_{att}\left(  \left\vert
\mathbf{r}_{1}\mathbf{-r}_{2}+\mathbf{K-L}+\mathbf{S}\right\vert \right)
d\mathbf{r}_{1}d\mathbf{r}_{2}%
\end{equation}
Because the attractive part of the potential typically varies slowly over the lengthscale of a hard sphere, it turns out to be quite accurate to
use the simple approximation $\widetilde{w}_{\mathbf{S}}^{\left(  \text{MF}\right) } = w_{att}(\mathbf{S})$ (see Supplementary text for illustration). In principle, one could approximate the potential by
some sort of interpolation - e.g. polynomial -  within each computational cell and evaluate the resulting expressions for the weights analytically but we find no advantage to this complication.
 Once again, once the weights $\widetilde
{w}_{\mathbf{S}}$ are known, the evaluation of the forces and energy is
efficiently coded via discrete Fourier transforms.

\subsection{Thermodynamics of the homogeneous fluid phases}

The thermodynamics of the bulk fluid phases of this model are simple to
describe. When the local density is constant, $\rho\left(  \mathbf{r}\right)
=\rho$, the fundamental measures are as well and have the values $\eta\left(
\mathbf{r}\right)  =\frac{\pi}{6}\rho d^{3}$, $s\left(  \mathbf{r}\right)
=\pi\rho d^{2}$ and $\mathbf{v}\left(  \mathbf{r}\right)  =0$. The free energy
functional becomes a simple function%
\begin{equation}
\frac{1}{V}\beta\Lambda\left(  \rho\right)  =\rho\ln\rho l^{3}-\rho+\rho
\frac{\eta\left(  4-3\eta\right)  }{\left(  1-\eta\right)  ^{2}}+\frac{1}%
{2}\beta a_{\text{vdw}}\rho^{2}-\widetilde{\mu}\rho
\end{equation}
where the Van der Waals constant is 
\begin{equation}
a_{\text{vdw}}=4\pi\int_{0}^{\infty}r^{2}w\left(  r\right)  dr\rightarrow\Delta^{3}%
\sum_{\mathbf{S}}\widetilde{w}_{\mathbf{S}}.
\end{equation}
The equilibrium state must minimize this so that the equilibrium density
$\overline{\rho}$ satisfies
\begin{equation}
\ln\overline{\rho}l^{3}+\frac{\overline{\eta}\left(  3\overline{\eta}%
^{2}-9\overline{\eta}+8\right)  }{\left(  1-\overline{\eta}\right)  ^{3}%
}+\beta a_{\text{vdw}}\overline{\rho}=\widetilde{\mu}%
\end{equation}
and the grand-canonical free energy is
\begin{equation}
\frac{1}{V}\beta\Omega=\frac{1}{V}\beta\Lambda\left(  \overline{\rho}\right)
=-\overline{\rho}\frac{\left(  1+\overline{\eta}+\overline{\eta}^{2}%
-\overline{\eta}^{3}\right)  }{\left(  1-\overline{\eta}\right)  ^{3}}%
-\frac{1}{2}\beta a_{\text{vdw}}\overline{\rho}^{2}=-\beta p
\end{equation}
where the last equality reminds us that this is just the negative of the
pressure. This can be written as a virial expansion,%
\begin{equation}
\frac{\beta p}{\rho}=\allowbreak1+\left(  \frac{2\pi d^{3}}{3}+\frac{1}%
{2}\beta a_{\text{vdw}}\right)  \rho+10\left(  \frac{\pi d^{3}}{6}\right)  ^{2}\rho
^{2}+18\left(  \frac{\pi d^{3}}{6}\right)  ^{3}\rho^{3}+...
\end{equation}
so that in particular, the second virial coefficient is $B=\frac{2\pi d^{3}%
}{3}+\frac{1}{2}\beta a_{\text{vdw}}$.

Two phase coexistence is possible when there are two solutions, $\overline
{\rho}_{1}$ and $\overline{\rho}_{2}$, which are both global minima of the
free energy function and so give equal free energies,%
\begin{align}
\ln\overline{\rho}_{1}l^{3}+\frac{\overline{\eta}_{1}\left(  3\overline{\eta
}_{1}^{2}-9\overline{\eta}_{1}+8\right)  }{\left(  1-\overline{\eta}%
_{1}\right)  ^{3}}+\beta a_{\text{vdw}}\overline{\rho}_{1}  &  =\ln\overline{\rho}_{2}%
l^{3}+\frac{\overline{\eta}_{2}\left(  3\overline{\eta}_{2}^{2}-9\overline
{\eta}_{2}+8\right)  }{\left(  1-\overline{\eta}_{2}\right)  ^{3}}+\beta
a_{\text{vdw}}\overline{\rho}_{2}\\
-\overline{\rho}_{1}\frac{\left(  1+\overline{\eta}_{1}+\overline{\eta}%
_{1}^{2}-\overline{\eta}_{1}^{3}\right)  }{\left(  1-\overline{\eta}%
_{1}\right)  ^{3}}-\frac{1}{2}\beta a_{\text{vdw}}\overline{\rho}_{1}^{2}  &
=-\overline{\rho}_{2}\frac{\left(  1+\overline{\eta}_{2}+\overline{\eta}%
_{2}^{2}-\overline{\eta}_{2}^{3}\right)  }{\left(  1-\overline{\eta}%
_{2}\right)  ^{3}}-\frac{1}{2}\beta a_{\text{vdw}}\overline{\rho}_{2}^{2}\nonumber
\end{align}
The spinodals are the densities at which the derivative of the pressure
vanishes,%
\begin{equation}
\left.  \frac{d\beta p}{d\rho}\right\vert _{s}=\frac{1+4\eta_{s}+4\eta_{s}%
^{2}-4\eta_{s}^{3}+\eta_{s}^{4}}{\left(  1-\eta_{s}\right)  ^{4}}+\beta
_{s}a_{\text{vdw}}\overline{\rho}_{s}=0
\end{equation}
which, incidentally, just requires solving a quintic polynomial equation to
determine the spinodal densities. The critical point occurs at the inflection
point of the pressure and so is determined by the system
\begin{align}
\left.  \frac{d\beta p}{d\rho}\right\vert _{c}  &  =\frac{1+4\eta_{c}%
+4\eta_{c}^{2}-4\eta_{c}^{3}+\eta_{c}^{4}}{\left(  1-\eta_{c}\right)  ^{4}%
}+\beta_{c}a_{\text{vdw}}\overline{\rho}_{c}=0\\
\left.  \frac{d^{2}\beta p}{d\rho^{2}}\right\vert _{c}  &  =\frac{\pi d^{3}%
}{6}\frac{4\left(  2+5\eta_{c}-\eta_{c}^{2}\right)  }{\left(  1-\eta
_{c}\right)  ^{5}}+\beta_{c}a_{\text{vdw}}=0\nonumber
\end{align}
Eliminating the mean-field term between these leaves an equation involving
only the density and this has only a single physical solution that is easily
determined numerically resulting in
\begin{align}
\rho_{c}  &  =\frac{0.249\,}{d^{3}}\\
k_{B}T_{c}  &  =-0.09\frac{a_{\text{vdw}}}{d^{3}}\nonumber\\
\frac{p_{c}}{\rho_{c}k_{B}T_{c}}  &  =0.357\nonumber
\end{align}

Note that the hard-sphere diameter is typically temperature-dependent so these
must be solved self-consistently. Nevertheless, it is interesting to observe
that the critical density is independent of the mean-field Van der Waals parameter $a_{\text{vdw}}$.

\section{Calculational Procedures}

\subsection{Gaussian profiles}

We have calculated results both by minimizing the discretized density field,
which we refer to as \textquotedblleft full\textquotedblright\ minimization.
We have also, for comparison, performed calculations by modeling the solid as
a sum of Gaussians at the lattice sites. For the latter, the density field is
\begin{equation}
\rho({\mathbf{r}})=(1-c)\left(  \frac{\alpha}{\pi}\right)  ^{3/2}\sum_{i}%
\exp\left(  -\alpha\left(  {\mathbf{r}}-{\mathbf{R}}_{i}\right)  ^{2}\right)
\end{equation}
where the sum is over the four positions of the molecules in the unit cell
($(0,0,0)$, $(a/2,a/2,0)$, $(0,a/2,a/2)$, $(a/2,0,a/2)$ where $a$ is the
lattice constant). The parameter $\alpha$ controls the width and $c$ is the
vacancy concentration (number of vacancies per lattice site). Although this
model can be implemented more efficiently, we have simply used it to determine
the discretized density from which the density functional is evaluated as in
the case of full minimization. The constrained, or Gaussian, minimization thus
consists of minimizing with respect to the two parameters $c$ and $\alpha$.
We carry out this minimization using the Nelder-Mead algorithm\cite{Nelder} as provided in the GSL library\cite{GSL}. 
Note that in any minimization of the FMT functional, it is possible that density fields are generated which cause the local packing fraction, $\eta({\bf r})$, to exceed one, which is outside the domain of the function. This simply indicates that the minimization routine has made too large of an adjustment to the density and in this case we return a large value for the free energy thus pushing the search back into the physical region.

\subsection{Full minimization}

For full minimization of the discretized density field
we start with an initial guess (based e.g. on minimized Gaussian profiles or the density field  minimized at some other thermodynamic parameters) and use the Fast Inertial
Relaxation Engine (FIRE) algorithm\cite{Fire}. This is a type of gradient
descent with inertia and we find it reliably converges in typically a few
thousand iterations for the case of the solid (and an order of magnitude
faster for the fluid). Details concerning the parameters used are given in the Supplementary text and here we only note that as in the case of the Gaussian profiles, care must be taken to backtrack if an attempted adjustment of the density takes it outside the physical domain. 

\subsection{Determining properties of the homogeneous solid phase}

A difficulty of the solid phase is that one must take explicit account of the
periodicity of the lattice, even when working directly with the density field.
This is because a homogeneous system is necessarily modeled using a finite
computational cell with periodic boundaries so that its size must be
commensurate with the lattice. An FCC solid can be described using a
non-primitive cubic simulation cell with lattice sites at each of the 8
corners as well as on the centers of each face. If the computational lattice
spacing is $\Delta$ and if the length of the cell is $N$ computational nodes,
then its physical length is $a = N\Delta$ which will then be the lattice
spacing of the solid phase. Our procedure is to fix these quantities and 
then minimize the density profile at constant chemical potentials $\mu_{0}$ as explained in the previous subsection.
The result is the free energy functional evaluated at these parameters, $\Omega(N, \mu_{0}; T, \Delta)$ and the corresponding density field. 
We then change the chemical potential to $\mu_{0} + \Delta\mu$ and use the previously determined density field as the initial guess for full minimization at this chemical potential. This is repeated over a
range of chemical potentials so that what we end up with is $\Omega(N, \mu
_{0}+m\Delta\mu; T, \Delta)$ for different values of $N$ and $m$. Then, for each value of $m$, i.e. for each chemical potential in the grid, we locate the three values of $N$ which
contain a minimum of the free energy and finally, estimate the the optimal (non-integer) value of $N$ by quadratic interpolation on these three values. This results in a list
of free energies, $\Omega(\mu_{0}+m\Delta\mu; T, \Delta)$, and these are used to find coexistence with the vapor and/or liquid phases (i.e. the values of chemical potentials where the two phases have equal free energies). This is again refined using quadratic interpolation. All quantities are then determined by the same
quadratic interpolation except the vacancy concentration which is sometimes
determined via linear interpolation (because the quadratic interpolation fails
for this quantity in some cases).

We note that an alternative would be to hold $N$ fixed and to vary the lattice
spacing $\Delta$. We choose not to to do this for several reasons. First, the
value of the VdW parameter, $a_{\text{VdW}}$ that determines the bulk thermodynamics
of the liquid phase changes as we change $\Delta$ but not when we change $N$
so that this introduces some unphysical variation into the model. Furthermore,
in this case, there are two ways to lower the density of the homogeneous
phase: by keeping the local density fixed and changing $\Delta$ or by holding
the spacing fixed and varying the local density. This makes the limit of the
homogeneous liquid ambiguous.

One might also wonder about the relevance of reporting a minimum in the
lattice spacing of fractional values of $N$ determined by interpolation. In
fact, the limitation to the discrete values of lattice parameter is an
artifact of trying to limit the cost of the calculations by using a minimal
cubic cell. If, e.g., the computational cell were two lattice spacings in
length, so that $2a = N \Delta$, then one could place one FCC lattice position
at the origin, $(0,0,0)$, one at $(N/2\Delta,0,0)$ and one at $(N\Delta, 0,0)$
so that the FCC lattice spacing would be $(N/2)\Delta$ allowing for
half-integer lattice constants. So, in principle, any rational lattice spacing
is possible, provided we use larger cells.

\subsection{Procedure to compute solid properties with the HCP lattice}

\begin{figure}[htp!]
	\includegraphics[trim={1.5cm 2.0cm 1.5cm 1.5cm}, clip=true,width=0.40\linewidth]{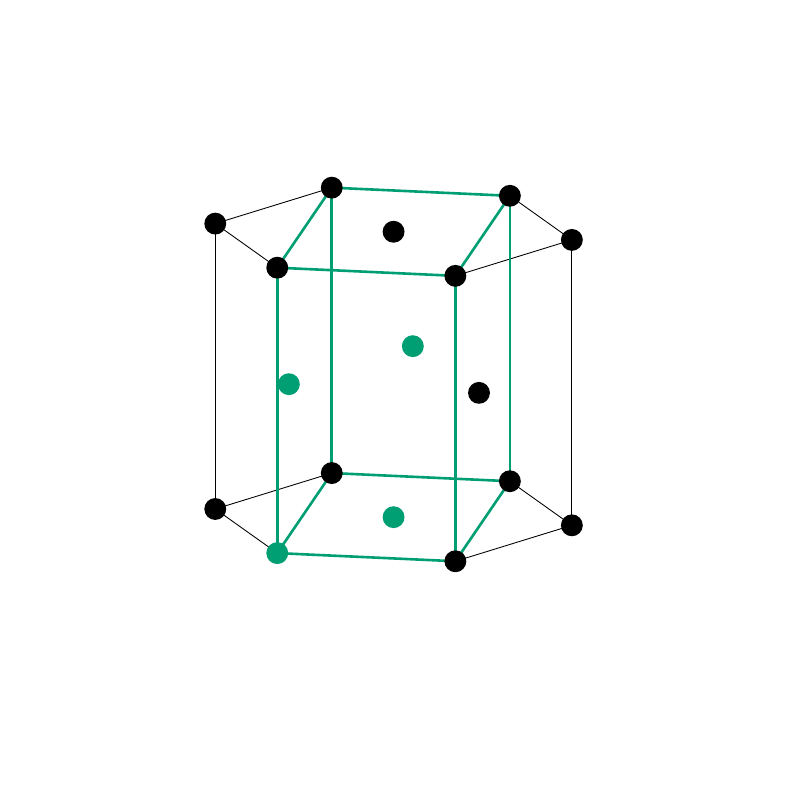}
	\caption{The HCP structure. The blue lines highlight the borders of the rectangular cell used for computations. The blue dots are the lattice sites which constitute the basis of the rectangular cell.}
	\label{HCPcell}
\end{figure}

In addition to the solid phase computations using the FCC lattice, we performed calculations for the HCP (hexagonal close-packed) lattice, using the Gaussian profiles. The geometry of the HCP structure makes it more difficult to study than the FCC one. This is because our implementation of classical DFT computations uses rectangular cells with the same grid spacing $\Delta$ in all directions. It is possible to use a rectangular cell to construct the HCP lattice that is compatible with periodic boundary conditions (see Fig. \ref{HCPcell}) but the ratio between the side's lengths are irrational numbers. That means we cannot construct the HCP lattice directly as our implementation requires all rectangular side's length to be multiples of the same grid spacing $\Delta$. Therefore, instead of directly constructing the HCP lattice, we compute the solid properties for regular hexagonal lattices near close-packing and interpolate the results for the ideal close-packing (HCP) proportions. 
More precisely, we first compute the rectangle side's lengths that are multiples of the grid spacing $\Delta$ and that are the closest to the ideal close-packing proportions.
They define our reference cell. Then, we generate
other rectangular cells by adding -1, 0 or 1 times the grid spacing to the side length of the reference cell,
for two of the three axis x, y, z. We perform computations for all of these cells and then interpolate the results successively along each of the two selected axis, using quadratic interpolations.

\section{Results}

We have performed calculations for three potentials:\ the Lennard-Jones
potential and two potentials recently proposed by\ Wang et al. The
Lennard-Jones potential is
\begin{equation}
v_{LJ}\left(  r\right)  =4\epsilon\left(  \left(  \frac{\sigma}{r}\right)
^{12}-\left(  \frac{\sigma}{r}\right)  ^{6}\right)
\end{equation}
which is then cutoff at a distance $r_{c}$ and shifted to give%
\begin{equation}
v_{LJ}\left(  r;r_{c}\right)  =\left(  v_{LJ}\left(  r\right)  -v_{LJ}\left(
r_{c}\right)  \right)  \Theta\left(  r_{c}-r\right)  .
\end{equation}
The LJ\ potential is widely used to model simple (e.g. atomic or
small-molecule)\ fluids and metals. The Wang-Ramirez-Hinestrosa, Dobnikar and
Frenkel (WHDF) potentials%
\begin{equation}
v_{WHDF}\left(  r\right)  =\alpha\left(  r_{c}\right)  \epsilon\left(  \left(
\frac{\sigma}{r}\right)  ^{2}-1\right)  \left(  \left(  \frac{r_{c}}%
{r}\right)  ^{2}-1\right)  ^{2}\Theta\left(  r_{c}-r\right)
\end{equation}
where%
\begin{equation}
\alpha\left(  r_{c}\right)  =2\left(  \frac{r_{c}}{\sigma}\right)  ^{2}\left(
\frac{3}{2\left(  \left(  \frac{r_{c}}{\sigma}\right)  ^{2}-1\right)
}\right)  ^{3}%
\end{equation}
This potential was introduced as a simplification relative to the LJ potential
which is designed to go smoothly to zero at its cutoff and to be deformable
between a LJ-like potential (when the cutoff is large, e.g. $r_{c}=2\sigma$)
and a colloid-like potential when the cutoff is small (e.g. $r_{c}=1.2\sigma
$). The difference in the two cases is attributable to the difference in the
width of the attractive well compared to the repulsive part of the potential
and results in qualitatively different phase diagrams as illustrated below.

\subsection{Surface tension}

\begin{figure}[htp!]
  \includegraphics[angle=0,scale=0.5]{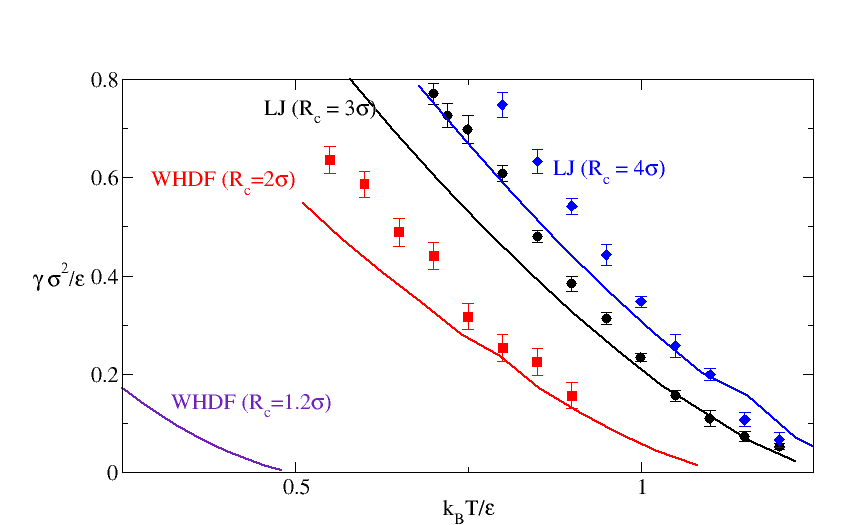}
\caption{Surface tension for the Lennard-Jones potential with cutoff $R_{c} = 3\sigma$ and $4\sigma$ and the WHDF potential for a simple fluid, cutoff of $R_{c} = 2\sigma$,  and for a colloidal system, $R_{c} = 1.2\sigma$. The calculations were performed using the DFT model with no adjustable parameters. Simulation results for the LJ system\cite{GrosfilsLutsko} and the WHDF systems\cite{Wang} are also shown with error bars of one standard error. }
\label{fig1}
\end{figure}

We have calculated the liquid-vapor surface tension for our model systems
using a cell consisting of a lattice of $1 \times1 \times N$ points with
$N=20,000$ and a spacing of $\Delta= 0.010$. For a given temperature, we first
determine the coexisting densities, $\rho_{\text{vap}}$ and $\rho_{\text{liq}%
}$ from the equation of state and the corresponding chemical potential. We
then create an initial density in which the two phases each occupy half the
computational cell with a sharp boundary between them. We minimize to get the
equilibrium density distribution $\rho^{\ast}(\mathbf{r})$ and compute the
surface tension as the excess surface free energy $\gamma= (\Omega[\rho^{\ast
}]-\Omega(\rho_{\text{vap}}))/(2\Delta^{2})$. The results are shown in Fig.
\ref{fig1} along with simulation results for the LJ system with the same
cutoff\cite{GrosfilsLutsko} and the results of Wang et al\cite{Wang}. The DFT
model, which has no adjustable parameters, compares reasonably well with simulation,
particularly for intermediate temperatures. Being a mean-field model, it is
not expected to capture the behavior near the critical point while the fact
that the mean-field contribution is motivated by the high-temperature limit in
liquid state theory\cite{HansenMcDonald} may account for the systematic
deviation at lower temperatures.

\subsection{Phase diagrams}

The vapor-liquid phase diagrams for our systems were already calculated in the
course of evaluating the surface tension. We have also calculated the
FCC-solid phase diagrams using a cubic cell with a fixed lattice spacing,
$\Delta$. The cell represents a non-primitive unit cell of the solid so that
its length is the lattice spacing of the FCC solid and in a perfect solid, the
total number of molecules in such a cell would be $4$. However, when
minimizing the free energy functional, the total number of molecules varies
and in general is less than this in the equilibrium state: the difference is a
measure of the equilibrium vacancy concentration in the solid phase which we
calculate as $c=\frac{4-N[\rho^{\ast}]}{4}$ where $N[\rho] = \int_{C}%
\rho(\mathbf{r}) d\mathbf{r}$ is the total number of molecules in the cell $C$.

\begin{figure}
[htp!]\includegraphics[angle=0,scale=0.45]{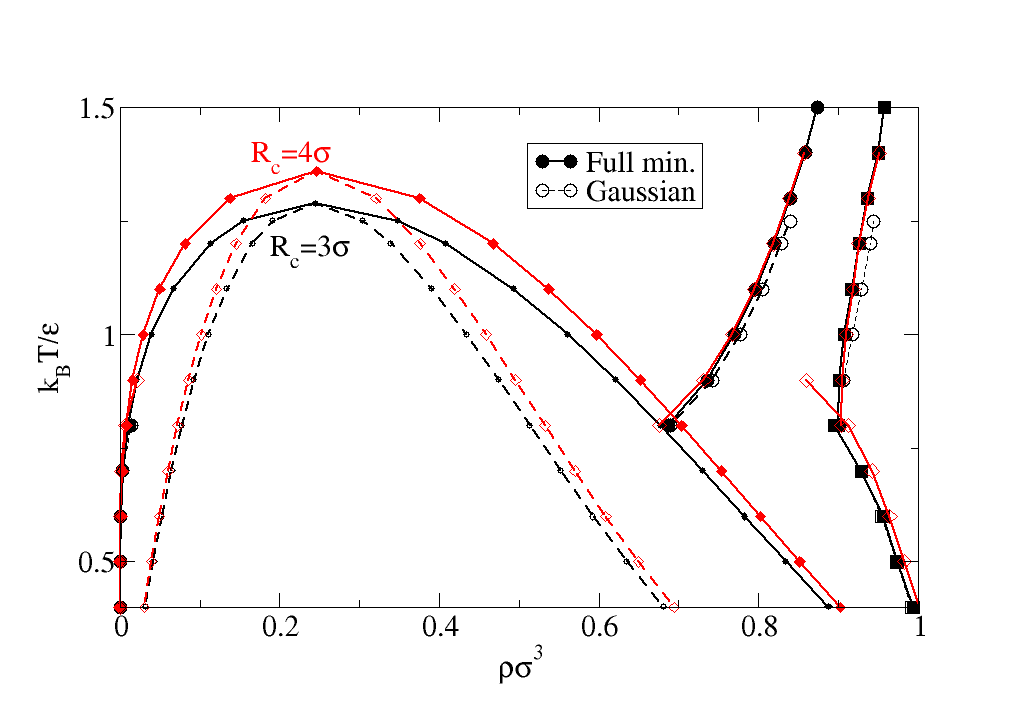}
\caption{Vapor-liquid-FCC phase diagram for the LJ potential for cutoffs of $3\sigma$ and $4\sigma$ as obtained by full minimization of the DFT functional. The broken lines show the spinodals. The results obtained using Gaussian profiles are also shown and are very close to the full minimization.}
\label{fig2}
\end{figure}

The resulting phase diagram for a Lennard-Jones potential is shown in Fig.\ref{fig2}. The mean-field model reproduces the correct qualitative behavior with vapor, liquid and solid binodals and a liquid-vapor spinodal. The Gaussian and full mininizations are in close agreement thus showing the close correspondence of the two. As seen in the Figure, the vapor-liquid critical point becomes lower as the cutoff of the potential is reduced but the quualitative behavior does not change. In contrast, the liquid-solid binodals show little sensitivity to the cutoff.

\begin{figure}[htp!]
  \includegraphics[angle=0,scale=0.45]{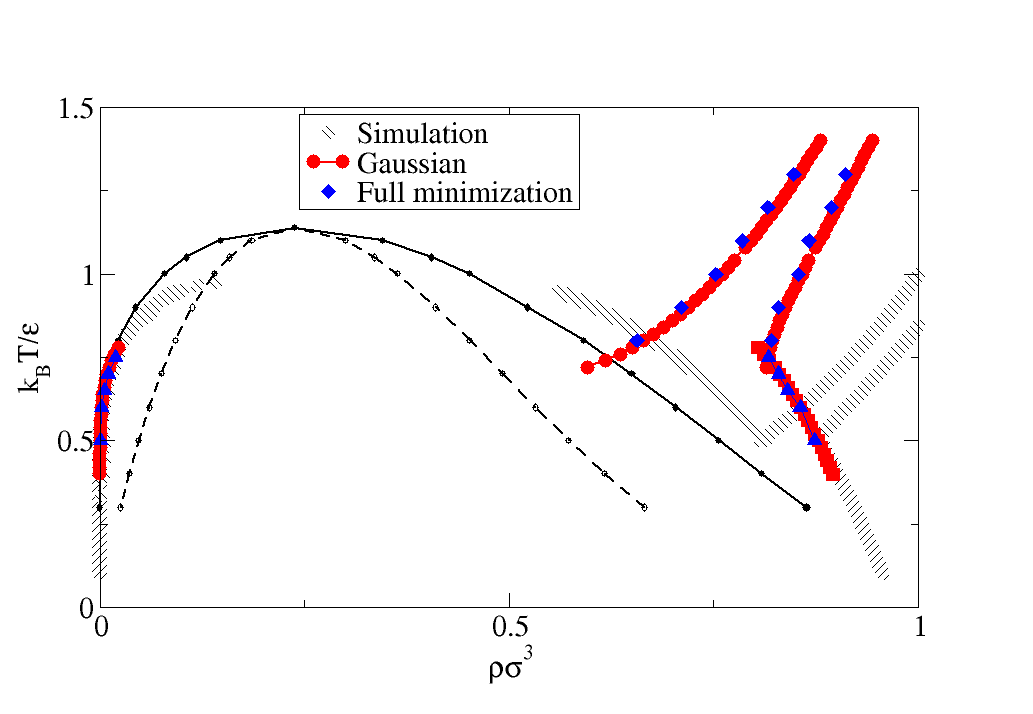}
\caption{Vapor-liquid-FCC phase diagram for the WHDF potential with range $R_{c} = 2\sigma$ which behaves like the Lennard-Jones potential as computed via full minimization of the DFT functional and using Gaussian profiles. The coexistence curves based on the data of Wang et al.\cite{Wang} are shown as shaded regions.}
\label{fig3}
\end{figure}

Figure \ref{fig3} shows the computed phase diagram for the WHDF potential with a cutoff of $2\sigma$ which is quite similar to the Lennard-Jones phase diagram. Comparison to the simulation data of Wang et al\cite{Wang} shows that, while qualitatively quite realistic, the mean-field model does not agree quantitatively with simulation. In particular, the fluid-solid binodals are displaced towards lower densities and higher temperatures than in the simulations. This is to be expected since no attempt is made in the mean-field model to reproduce even the fluid-phase thermodynamics quantitatively.

\begin{figure}[htp!]
  \includegraphics[angle=0,scale=0.45]{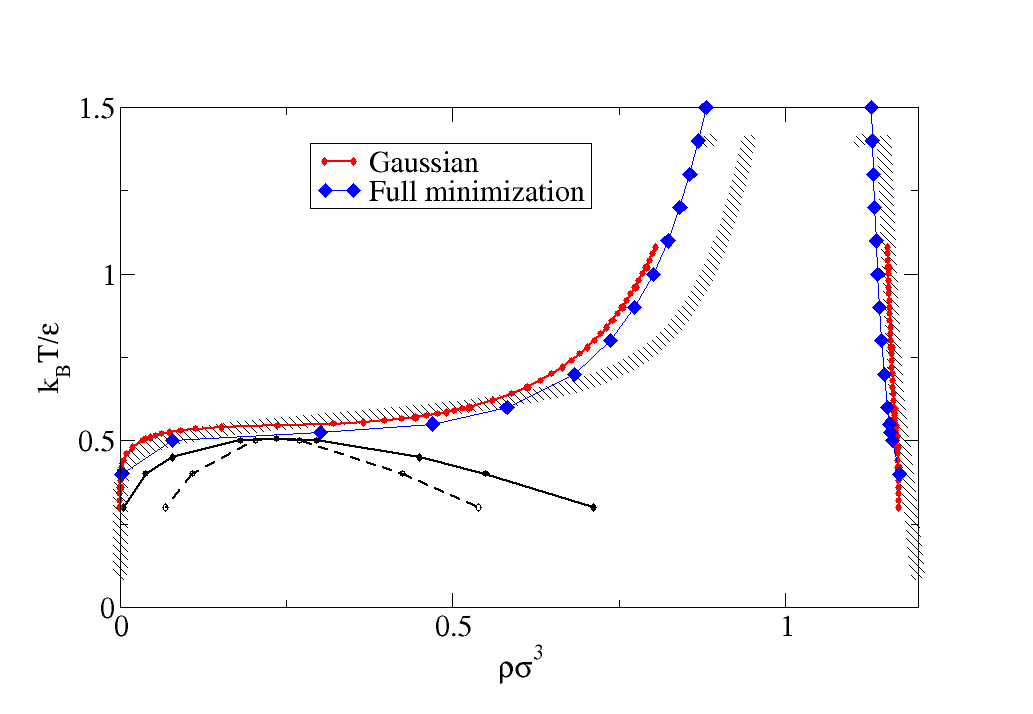}
\caption{Vapor-liquid-FCC phase diagram for the WHDF potential with range $R_{c} = 1.2\sigma$ which models a colloidal phase diagram  as computed via full minimization of the DFT functional and using Gaussian profiles. The coexistence curves based on the data of Wang et al.\cite{Wang} are shown as shaded regions.}
\label{fig4}
\end{figure}

The results for the WHDF potential with a smaller cutoff, producing a colloid-like phase diagram, are shown in Fig.\ref{fig4} where the typical suppression of the liquid-vapor transition into the metastable region of the fluid-solid transition is evident. The cDFT again faithfully reproduces this qualitative behavior and, indeed, is in reasonable quantitative agreement with the simulations. In summary, while the cDFT is not reliably quantitatively accurate, it does a good job of tracking the qualitative behavior resulting from variations in the interaction potential.

\bigskip

\subsection{Vacancies}

\begin{figure}
[htp!]\includegraphics[angle=0,scale=0.5]{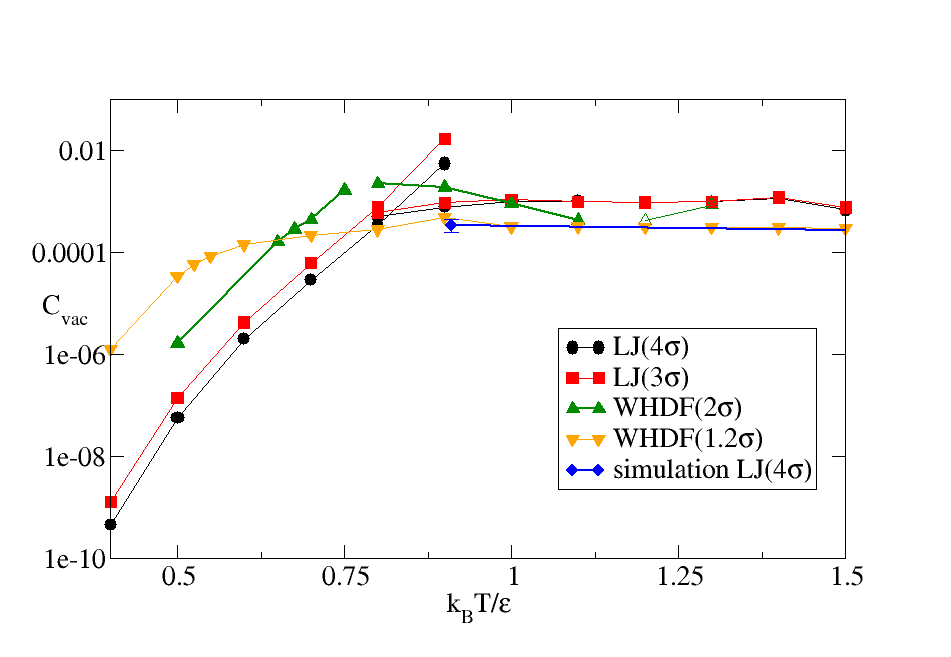}
\caption{The vacancy concentration for the colloidal potential WHDF with cutoff $1.2\sigma$ on the solid-fluid coexistence curves and the simple fluid potentials on the liquid-solid (higher temperature) and vapor-solid (lower temperature) coexistence curves,  obtained with the full minimization. In the case of the WHDF potential with cutoff $2\sigma$, the open symbols are values for which $c$ is negative and so the absolute value is shown on the logarithmic scale used here.}
\label{fig5}
\end{figure}

\begin{figure}[htp!]
	\includegraphics[width=0.80\linewidth]{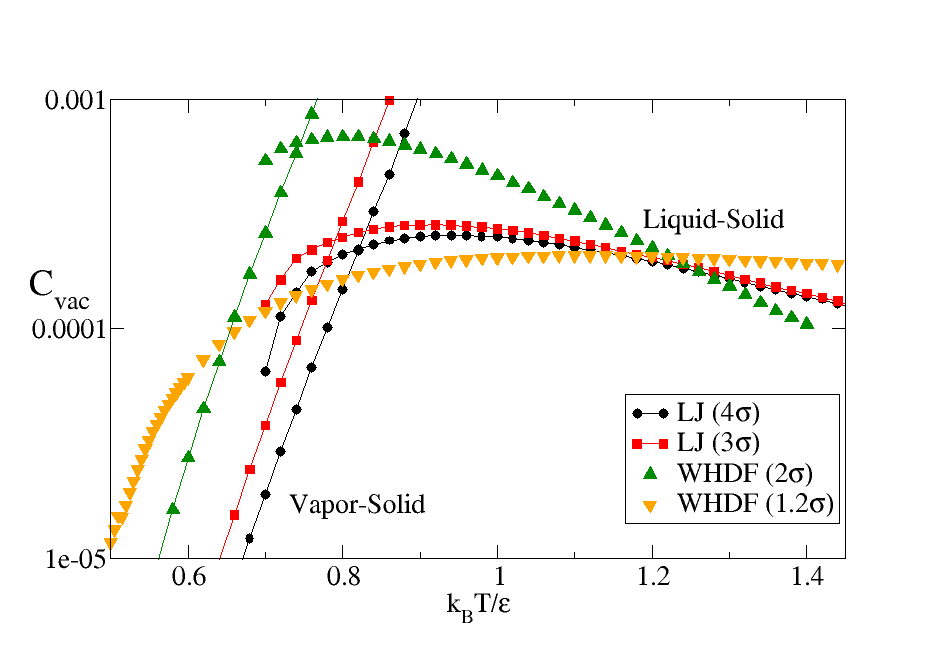}
	\caption{The same as Fig. \ref{fig5} but showing the results in the Gaussian approximation}
	\label{fig6}
\end{figure}

\begin{figure}[htp!]
	\includegraphics[width=0.80\linewidth]{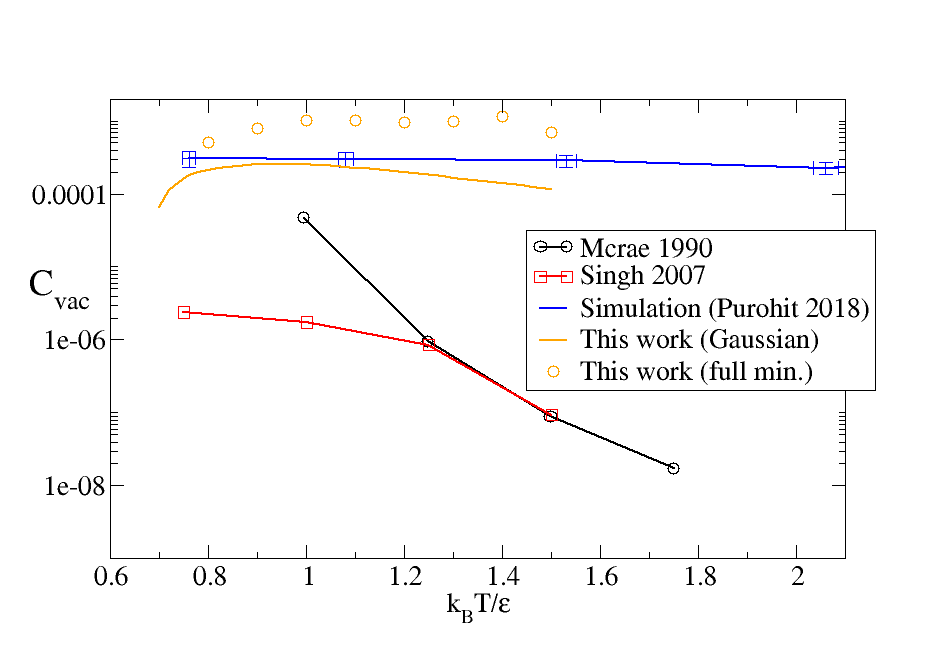}
	\caption{Comparison of vacancy concentrations on the solid-liquid coexistence curve for the Lennard-Jones potential with cutoff $4\sigma$. This graph shows our results (line and open circles) next to simulations\cite{Purohit} (line with error bars) and the older calculations of Singh et al\cite{Singh} and Mcrae\cite{Mcrae}. }
	\label{fig7}
\end{figure}

The vacancy concentration determined by the DFT calculations is shown as a
function of temperature in Fig. \ref{fig5} for full minimization and in Fig. \ref{fig6} in the Gaussian approximation. The results are qualitatively the same in the two cases and the numerical differences, while real, are modest overall. In Fig. \ref{fig7} we compare both calculations to some simulation
data for the LJ potential, but results are only available for higher
temperatures. For the LJ system, the vacancy concentration is relatively
insensitive to the cutoff and and is nearly constant for liquid-solid
coexistence, at least in the range of temperatures reported here. This is
qualitatively consistent with the simulation results and is much better than
older calculations that used more primitive models of the hard-sphere free
energy functional\cite{Mcrae, Singh}. At lower temperatures, on the solid-vapor
coexistence curve, it drops sharply with temperature as one would expect. The
WHDF potential for simple fluids gives similar results near the triple point
but drops as the temperature increases and actually becomes negative
(indicating interstitials rather than vacancies). This may well be an
unphysical artifact of the cDFT calculations - or due to inaccuracies
in the interpolations - but there are no simulation results to compare to. We note that the Gaussian minimizations do not show this behavior.
Finally, the results for the WHDF colloidal potential seem reasonable but
there are again no independent results for comparison.

\subsection{Density profiles}

\begin{figure}
[htp!]\includegraphics[angle=0,scale=0.5]{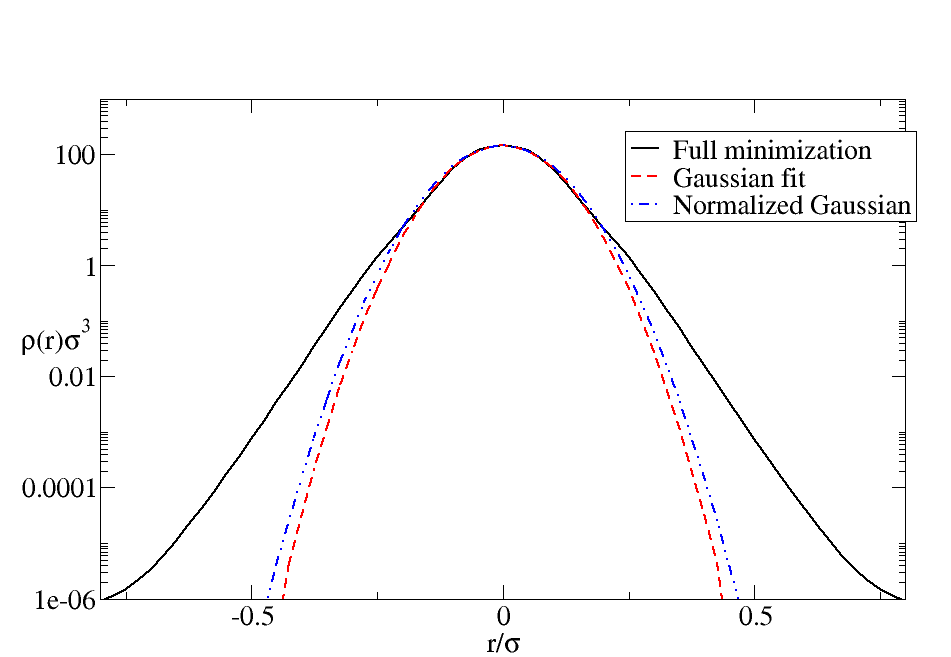}
\caption{Density distribution near a lattice site for a LJ solid with cutoff $3\sigma$, $k_{B}T = 0.8$, $N = 66$ lattice points in each direction and $\beta \mu = -3$ which is near the triple point. Also shown are a best fit to a Gaussian (broken line) and to a normalized Gaussian (dash-dot line). The widths for the latter two fits are $\alpha \approx 95$ and $86$, respectively. Significant variation from the Gaussians in the tail region cannot be accounted for by nearest-neighbor contributions.}
\label{fig9}
\end{figure}

\begin{figure}
[htp!]\includegraphics[angle=0,scale=0.5]{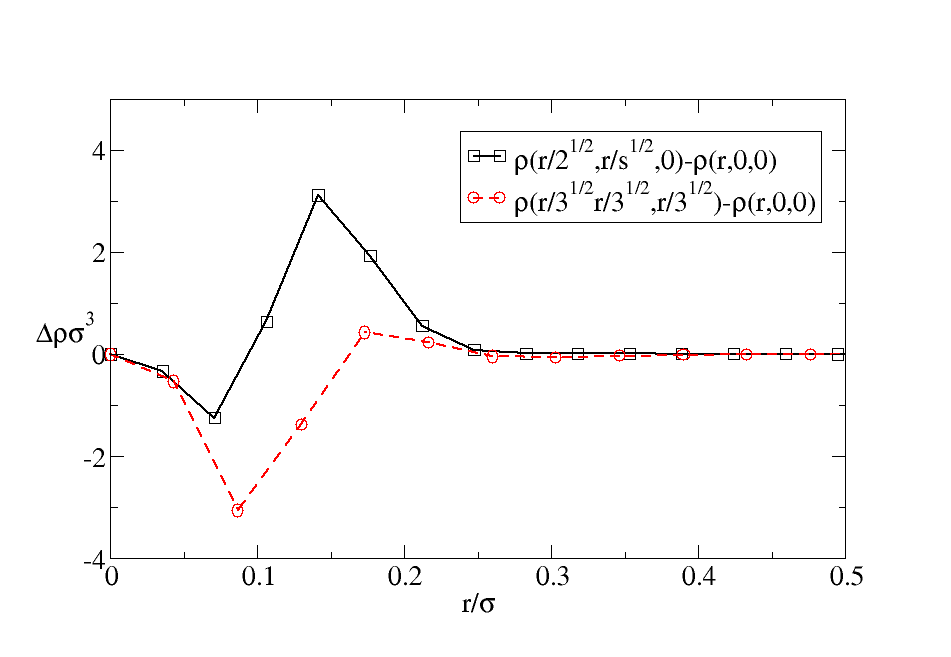}
\caption{Illustration of asymmetries in the density distribution. The plot shows the difference in density near a lattice site moving along the $(110)$ (black, full line and squares) and $(111)$ (red, broken line and circles) directions compared to that along the $(100)$ direction. The conditions are the same as for Fig. \ref{fig6}.}
\label{fig10}
\end{figure}

It has been seen that the use of Gaussian profiles produces results which are
quantitatively very similar to the results of unconstrained minimization of
the cDFT functional. Nevertheless, differences are expected due to the fact
that the neighborhood of a lattice position in the solid is not spherically
symmetric so that the Gaussian profiles, which force spherical symmetry of the
contribution of each lattice site, can only be an approximation. In Fig.
\ref{fig9} we show the density profile for a LJ solid near the triple point
($k_{B}T=0.8$, $a = 66 \times0.025\sigma= 1.65\sigma$). The central region of
the profile is well-fit by a Gaussian but away from the center, the
distribution is much broader than a Gaussian. The difference is not due to
contributions from the neighbors: the nearest neighbor distance is
$a/\sqrt 2$ and the best-fit Gaussian is $145 e^{-95r^{2}}$ and a best fit
normalized Gaussian has  $\alpha= 86$. Even taking the slightly broader
latter function, the contribution at the half the nearest neighbor distance is
$\left(  \frac{86}{\pi} \right)  ^{1.5}e^{-86( 0.5 \times a/\sqrt 2)^{2}}
\approx3 \times10^{-11}$ which is much smaller than the excess in the tail of
the density. Figure \ref{fig10} shows the difference in density along lines
running from a lattice position in the direction $(110)$ and $(111)$ relative
to that along the $(100)$ direction and one sees that at intermediate
distances, there is some asymmetry. This may be attributable to the fact that
moving along the $(110)$ direction means towards a nearest neighbor and at
this density, the nearest neighbor distance, $1.167\sigma$ is slightly more
than the position of the minimum of the potential well, $1.12\sigma$, giving a
small preference for adding density in that direction at the expense of the others.

\subsection{Relative stability of FCC and HCP configurations}

\begin{figure}[htp!]
	\includegraphics[width=0.70\linewidth]{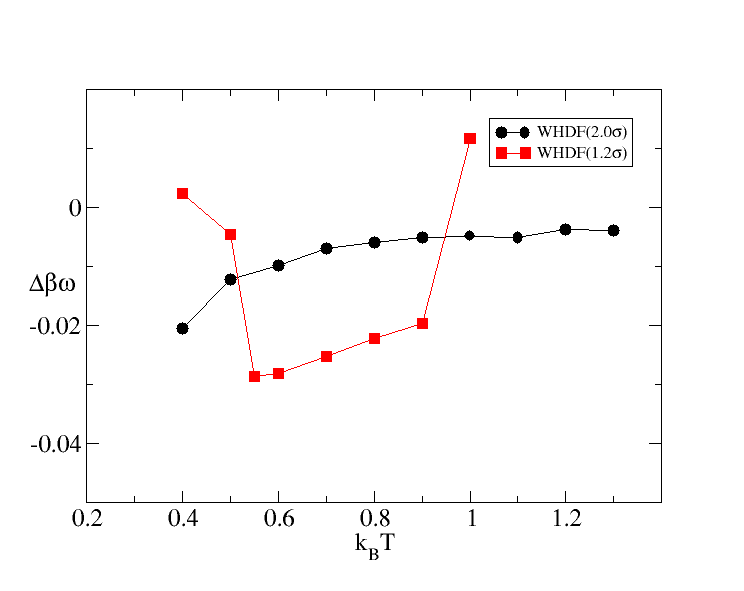}
	\caption{Free energy difference per unit volume, $\beta \omega = \beta \omega_{\text{HCP}}-\beta \omega_{\text{FCC}}$ as a function of temperature computed using Gaussian profiles and the WHDF potential. For each temperature, the computations of both FCC and HCP phases were performed at the same chemical potential, close to the corresponding fluid-solid coexistence. We estimate that the numerical errors in the differences are on the order of $1 \times 10^{-3}$  for the larger cutoff and $1 \times 10^{-2}$  for the smaller one.}
	\label{FCCvsHCP}
\end{figure}

The free energy difference between the FCC and HCP configurations is reported on Fig.\ref{FCCvsHCP} as a function of the temperature. These DFT computations use the Gaussian profiles to parameterize the density field and have been performed for the WHDF potential with $r_c=1.2$ (colloid-like) and $r_c=2.0$ (simple fluid). For both potentials we get very similar free energies for the two configurations, about the order of magnitude of the expected interpolation error which is estimated to be  $\Delta \beta\omega \sim 10^{-3}$ for the  larger cutoff and an order of magnitude larger for the smaller cutoff. Such small differences are expected because these two structures only differ at the second neighbor. Wang et al. also state that simulations for the WHDF potential lead to very similar free energies for the FCC and HCP structures \cite{Wang}. Our results indicate that the presented DFT model correctly reproduce this behavior and that using the FCC lattice gives a reasonable description of the solid free energies whether or not it is the most stable configuration.

\color{black}

\section{Conclusions}
We have presented a fully robust finite-elements implementation of the standard cDFT model. The novel elements of our implementation are
(a) it is entirely formulated in real-space and only uses discrete FFT's to evaluate discrete convolutions efficiently rather than mixing
discrete real-space quantities and analytically Fourier-transformed continuous quantities in an uncontrolled manner and (b) the real-space
weights needed in FMT are evaluated analytically thus avoiding the need for using tables of pre-compiled spherical integration points as were
previously used\cite{LutskoLam}. The latter fact significantly reduces numerical noise in the calculations and, together with the intrinsic stability of the methods, we routinely do minimizations of the solid phase at constant chemical potential which, as noted in the Introduction, has previously proven infeasible.

We have used this algorithm to first
reproduce standard results for the Lennard-Jones potential and then to test the robustness of cDFT in describing more general phase diagrams as
result from the WHDF potentials. We confirm that the model performs well across all tests. We believe that this implementation is well suited
for more challenging applications such as dynamic DFT and studies of nucleation. Finally, we have verified the accuracy of the computationally
cheap Gaussian approximation which opens the possibility to the use of pseudo-spectral methods with radial basis functions that could provide a
more efficient alternative to the finite-difference algorithms commonly used.

\bibliography{solid}

\appendix

\section{The FMT weights \label{AppW}}

In this Appendix, we give a somewhat simplified form for the analytic determination of the discrete FMT weights which is discussed in detail in the Supplementay material.
For the local packing fraction, one needs to evaluate%
\begin{equation}
w^{\left(  \eta\right)  }\left(  \mathbf{S}\right)  =\sum_{\mathbf{I}=0,1}%
\int_{\mathcal{C}\left(  \mathbf{0}\right)  }\Theta\left(  R-\left\vert
\mathbf{S}+\mathbf{I}\right\vert \right)  A_{\mathbf{I}}\left(  \mathbf{r}%
\right)  d\mathbf{r.}%
\end{equation}
Before tackling this, we note that the other fundamental measures follow
directly from this one via%
\begin{align}
w^{\left(  s\right)  }\left(  \mathbf{S}\right)   &  =\frac{\partial}{\partial
R}w^{\left(  \eta\right)  }\left(  \mathbf{S}\right)  \\
w^{\left(  \mathbf{v}\right)  }\left(  \mathbf{S}\right)   &  =-\frac
{\partial}{\partial\mathbf{S}}w^{\left(  \eta\right)  }\left(  \mathbf{S}%
\right)  \nonumber
\end{align}
so we need only concentrate on determining the first one. The details are given in the Supplementary material and here we just report the result which has been somewhat simplifed. In
general, for all of the weights one finds that 
\begin{align}
\widetilde{w}^{\left(  \alpha\right)  }\left(  \mathbf{S}\right)
=\sum_{\mathbf{I}\in\left\{  -1,1\right\}  }\sum_{\mathbf{K}=\mathbf{0}%
,\mathbf{I}}\Theta\left(  R^{2}-\left(  S_{x}+K_{x}\right)  ^{2}-\left(
S_{x}+K_{y}\right)  ^{2}-\left(  S_{x}+K_{z}\right)  ^{2}\right)  \left(
-1\right)  ^{K_{x}+K_{y}+K_{z}} \\
\times G^{\left(  \alpha\right)  }\left(
\mathbf{S+I,S+K}\right)  . \nonumber
\end{align}
For the local packing fraction, and for the positive octant $S_{j}\geq0$, the required function is
\begin{align}
G^{\left(  \eta\right)  }\left(  \mathbf{T,V}\right)   &  =-\frac{1}%
{48}\left(  R^{2}-V^{2}\right)  ^{3}-\frac{1}{8}\left(  R^{2}-V^{2}\right)
^{2}\left(  \mathbf{T\cdot V}\right)  \\
&  -T_{x}T_{y}T_{z}V_{x}V_{y}V_{z}+\frac{\pi}{6}T_{x}T_{y}T_{z}R^{3}%
+\mathcal{P}_{xyz}J\left(  \mathbf{T,V}\right)  \nonumber
\end{align}
with%
\begin{align}
J\left(  \mathbf{T,V}\right)   &  =-\frac{1}{6}T_{x}T_{y}T_{z}R^{3}%
\arcsin\frac{V_{x}V_{y}}{\sqrt{\left(  R^{2}-V_{x}^{2}\right)  \left(
R^{2}-V_{y}^{2}\right)  }}\\
&  +\frac{1}{24}T_{y}T_{z}\left(  3\left(  R^{2}-V_{x}^{2}\right)  ^{2}%
+4T_{x}V_{x}\left(  3R^{2}-V_{x}^{2}\right)  \right)  \left(  \allowbreak
\arcsin\frac{V_{y}}{\sqrt{R^{2}-V_{x}^{2}}}-\frac{\pi}{4}\right)  \nonumber\\
&  +\frac{1}{120}T_{z}\left(  4\left(  R^{2}-V_{x}^{2}-V_{y}^{2}\right)
^{2}+5T_{x}V_{x}\left(  5R^{2}-2V_{x}^{2}-5V_{y}^{2}\right)  +20T_{x}%
T_{y}V_{x}V_{y}\right)  \sqrt{R^{2}-V_{x}^{2}-V_{y}^{2}}\nonumber\\
&  -\frac{1}{12}\left(  T_{x}V_{x}^{3}+3T_{x}T_{y}V_{x}V_{y}\right)  \left(
\allowbreak\allowbreak R^{2}-V^{2}\right)  -\frac{1}{30}T_{x}V_{x}^{5}%
-\frac{1}{3}T_{x}T_{y}V_{x}^{3}V_{y}\nonumber
\end{align}
where $\mathcal{P}_{xyz}$ is an operator indicating a sum over all 6
permutations of the x,y and z components of the vectors (performed
simultaneously:\ that is, one element of the sum is $J\left(  T_{x}%
,T_{y},T_{z},V_{x},V_{y},V_{z}\right)  $, another is $J\left(  T_{y}%
,T_{x},T_{z},V_{y},V_{x},V_{z}\right)  $, etc.). For the packing fraction, the
spherical symmetry implies that the weights for other octants (e.g. with
$S_{x}<0)$ are the same as for $\left\vert S_{x}\right\vert $. The weights for
the other fundamental measures follow from this result via differentiation
(they are given explicitly in the Supplementary text). Note that it is easy to
see that $J\left(  \mathbf{T,V}\right)  =0$ if $V=R$ so terms involving a
derivative of the step function gives no contribution.   

These results have been checked by (a) independently deriving the result for
$w^{\left(  s\right)  }$and verifying that it agrees with that derived by
differentiation ); (b) by evaluating the three-dimensional integrals
numerically and comparing to the analytic result given here; and (c)\ by
comparing the resulting free energies to those calculated using other
implementations. All tests confirm the validity of the present results.

\end{document}